\documentclass[journal]{IEEEtran}
\usepackage{amsfonts}
\usepackage{amsmath}
\usepackage{amssymb}
\usepackage{array}
\usepackage{mathrsfs}
\usepackage{leftidx}
\usepackage{graphicx}
\usepackage{dcolumn}
\usepackage{bm}
\usepackage[usenames]{color}
\usepackage{colortbl,booktabs}
\usepackage{subfigure}
\usepackage{cite}
\usepackage{multirow}

\newcommand{\tr}{{\rm Tr}}%

\begin{document}
%
\title{Minimum-Time Selective Control of Homonuclear Spins}


\author{Tian-Ming Zhang,
Re-Bing Wu,~\IEEEmembership{Member,~IEEE},
Fei-Hao Zhang,
Tzyh-Jong Tarn,~\IEEEmembership{Life~Fellow,~IEEE}, \\
and Gui-Lu Long
\thanks{This work was supported by TNlist and National Natural Science Foundation of China (Grant Nos. 61374091, 61134008, 11175094 and 91221205), the National Basic Research Program of China under Grant No.2011CB9216002. GLL also thanks the support of Center of Atomic and Molecular Nanoscience of Tsinghua University.}
\thanks{Tian-Ming Zhang and  Re-Bing Wu are with the Department of Automation,
Tsinghua University, and also with the Center for
Quantum Information Science and Technology (TNlist), Beijing 100084, China
(e-mail: rbwu@tsinghua.edu.cn).}
\thanks{Fei-Hao Zhang and Gui-Lu Long are with Department of Physics, Tsinghua University,
Collaborative Innovation Center of Quantum Matter, and the Center for
Quantum Information Science and Technology (TNlist), Beijing 100084, China
(e-mail: gllong@tsinghua.edu.cn).}
\thanks{Tzyh-Jong Tarn is with Department of Electrical and Systems Engineering,
Washington University, St. Louis, MO 63130, USA. He is also with the Center for
Quantum Information Science and Technology (TNlist), Beijing 100084.}
}


\IEEEtitleabstractindextext{%
\begin{abstract}
In NMR (Nuclear Magnetic Resonance) quantum computation, the selective control of multiple homonuclear spins is usually slow because their resonance frequencies are very close to each other. To quickly implement controls against decoherence effects, this paper presents an efficient numerical algorithm for designing minimum-time local transformations in two homonuclear spins. We obtain an accurate minimum-time estimation via geometric analysis on the two-timescale decomposition of the dynamics. Such estimation narrows down the range of search for the minimum-time control with a gradient-type optimization algorithm. Numerical simulations show that this method can remarkably reduce the search efforts, especially when the frequency difference is very small and the control field is high. Its effectiveness is further demonstrated by NMR experiments with two homunuclear carbon spins in a trichloroethylene ($\rm C_2H_1Cl_3$) sample system.
\end{abstract}

}
\maketitle
\IEEEdisplaynontitleabstractindextext
\IEEEpeerreviewmaketitle

\section{Introduction}\label{Sec:Introduction}
\IEEEPARstart{H}{o}monuclear spin systems are referred to as molecular systems that contain nuclear spins with the same type of natural atoms (e.g., carbon spins in the same molecule) which are prevalent in NMR (nuclear magnetic resonance) based quantum information processing with many qubits \cite{ernst1987principles,levitt2013spin,vandersypen2005nmr,steffen2001toward,negrevergne2006benchmarking} because only a finite number of spin-1/2 nuclear spins can be encoded as qubits. Unlike heteronuclear spins that can be individually addressed by resonant magnetic fields, selective control of homonuclear spins is much harder due to their tiny differences between each other. With a common magnetic control field, their motions are usually discernable after a long time, which is unwanted because decoherence may gradually destroy quantum coherence in the spins. In order to accelerate the operation, the control sequence can be selected as a solution to the minimum-time control problem that has been extensively studied in the literature. For single spin systems, the minimum-time gate control solution was obtained \cite{boozer2012time,Wu2002,Boscain2005, boscain2006time} via Pontryagin Minimum Principle \cite{Bryson1975}. For two or three heteronuclear spins (i.e., nuclear spins of difference types of atoms), the Cartan decomposition of the controllability Lie algebra was used to calculate the minimal time required for quantum transformations under hard pulses \cite{Khaneja2001,Khaneja2002}. For systems with bounded controls, the determination of minimum-time quantum evolution can also be formulated as a quantum brachistochrone problem \cite{Carlini2006,Carlini2007,Carlini2008,Carlini2012time}. In Ref.\cite{tibbetts2012exploring}, the Pareto front was explored for understanding the trade-off between the competitive objectives of maximizing the transformation fidelity and minimizing the control time.

The control of homonuclear spin systems is closely related to the optimal dynamical discrimination (ODD) of molecular control systems \cite{Li2002,Petersen2010}, both aiming at manipulating dynamically similar systems. In Ref.~\cite{d2002optimal,d2003controllability}, the form of optimal control and the underlying controllability Lie algebra structure are analyzed for two homonuclear spin systems, and geometric analyses show that the optimal trajectories can be selected among singular extremal solutions of the Pontryagin Maximum Principle~\cite{bonnard2012geometric}. In particular, it was found that the minimum-time control for simultaneous inversion of two homonuclear spins is bang-bang~\cite{assemat2010simultaneous}. Other problems, such as the maximization of signal to noise ratio, was also investigated for two-spin cases~\cite{pry2006optimal}.

The optimization of homonuclear spin systems generally does not have analytical solutions except in rare cases, and numerical algorithms are needed for the optimization. In the literature, gradient-based algorithms have been successfully applied to quantum optimal control problems \cite{Khaneja2005,Schirmer2009,Machnes2011,Rowland2012}, among which many were realized in NMR system as a good testbed for quantum control \cite{Ryan2008}. However, since the final time $T$ is fixed, an iterative strategy need to be designed to numerically locate the minimum time. For example, in \cite{lapert2012time} a monotonically convergent algorithm is proposed to simultaneously minimize the time. As will be seen below, we will make use of the multi-timescale property of homonuclear spin dynamics to estimate with high precision the the minimal time duration, according to which the search efforts with any numerical algorithm may be greatly reduced.

The paper is organized as follows. Section~\ref{Sec:Analysis} provides the control model for homonuclear spin systems in NMR experiments, based on which a minimum-time estimation formula is presented based on a two timescale geometric analysis. Section~\ref{Sec:SimExpRes} introduces the numerical algorithm for seeking minimum-time controls of local transformations based on the estimated minimum time, whose effectiveness is demonstrated by numerical simulations in Section~\ref{Sec:SimRes} and experiments in Section~\ref{Sec:ExpRes}. Finally, Section~\ref{Sec:Conclusion} concludes the results.

\section{Geometric analysis for minimum-time design}\label{Sec:Analysis}
This section will summarize the model for multiple homonuclear spin systems, following which an estimation formula will be derived for the minimum time required for two-spin local transformations.

\subsection{Control system model}
Consider a quantum homonuclear system that contains $N$ homonuclear spins. The dynamics is governed by the following Schr\"{o}dinger equation
\begin{equation} \label{eq:ControlSystemModel}
  i \dot U(t)=H_{\rm tot} (t)U(t),
\end{equation}
where the evolution operator $U(t)$ is a $2^N$-dimensional unitary matrix. The control of these spins is through a radiofrequency (RF) magnetic field whose carrier frequency is $\omega_{\rm rf}$. The total Hamiltonian $H_{\rm tot}(t)$ in the rotating frame (with angular frequency $\omega_{\rm rf}$) consists of the following three parts\cite{vandersypen2005nmr,altafini2012modeling}:
\begin{eqnarray} \label{eq:SystemModelHam}
 \label{eq:ZeemanHam} H_{\rm Z} &=&  -\sum_{k=1}^N \left[ (1-\delta_k)\omega_0-\omega_{\rm rf}\right]S^k_z,  \\
 \label{eq:JcouplingHam} H_{J}
&=&\sum_{1\leq i<j\leq N} 2\pi   J_{ij}(S_x^i S_x^j+S_y^i S_y^j+S_z^i S_z^j), \\
 \label{eq:ControlHam}H_{\rm RF} &=&  -\sum_{k=1}^N(1-\delta_k)\left[\omega_x(t)S^k_x +\omega_y(t) S^k_y\right],
\end{eqnarray}
where $S_\alpha^k=I_2^{\otimes(k-1)}\otimes{\sigma_\alpha}\otimes I_2^{\otimes(N-k)}$ with $\otimes$ being the Kronecker product and $\alpha=x,y,z$. Here, $I_2$ is the two-dimensional identity matrix and
$$\sigma_x=\frac{1}{2}\left[\begin{array}{cc}0&1\\1&0\end{array}\right],
~~\sigma_y=\frac{1}{2}\left[\begin{array}{cc}0&-i\\i&0\end{array}\right],
~~\sigma_z=\frac{1}{2}\left[\begin{array}{cc}1&0\\0&-1\end{array}\right]$$
are Pauli matrices.

The Hamiltonian $H_{\rm Z}$ characterizes the Zeeman splitting by a strong static magnetic field in $z$-axis, in which $\omega_0 = \gamma B_0$ with $\gamma$ being the gyromagnetic ratio of the nuclear spin and $B_0$ being the strength of the static magnetic field. The Larmor frequency $(1-\delta_k)\omega_0$ of each homonuclear spin is slightly different from $\omega_0$ by the chemical shift $\delta_k \ll 1$ induced by its environment.

The weak and isotropic $J$-coupling Hamiltonian $H_J$ comes from the indirect electron-mediated interaction. The values of coupling constants $J_{ij}$ between spins $i$ and $j$ range from a few hundred Hertz for one-bond couplings to only a few Hertz for three- or four-bond couplings.

The control Hamiltonian $H_{\rm RF}$ is invoked by a radiofrequency magnetic field whose intensities in $x$ and $y$ axes are $\omega_{x}(t)$ and $\omega_{y}(t)$, respectively. The effective action on each spin is also differentiated by the chemical shifts. Due to the power limitation, the control field  are subject to the following bound constraint
\begin{equation}  \label{eq:ConstrainedControl}
\omega_x^2(t) + \omega_y^2(t) \leq \Omega^2,
\end{equation}
where the bound $\Omega$ is determined by the maximum power available in the NMR spectrometer.

\subsection{Estimation of minimum time in two-spin systems}\label{Sec:MinimumControlForTwoQubit}
Our goal is to find the shortest time duration and corresponding control functions $\omega_x(t)$ and $\omega_y(t)$ that steer the propagator $U(T)$ to a target transformation $U_f\in [{\rm SU}(2)]^{\otimes N}$ under the constraint (\ref{eq:ConstrainedControl}). Such transformation represents a local operation on the spins, e.g., the following transformation
$$U_f=R_{x,y,z}(\theta_1)\otimes R_{x,y,z}(\theta_2)\!=\!e^{-i \theta_1\sigma_{x,y,z}}\otimes e^{-i\theta_2\sigma_{x,y,z}}$$ simultaneously rotates two spins around $x$ (or $y,z$) axis by $\theta_1$ and $\theta_2$, respectively.

To facilitate the estimation of the minimum time for local transformations, it is reasonable to omit $H_J$ because the frequency difference between homonuclear spins are usually much greater (up to two to three orders of magnitude) than the $J$-coupling parameters. Thus, the overall unitary propagator is approximated as
\begin{equation} \label{eq:WholeUnitaryOtimeFrom}
U(t)\approx U_1(t)\otimes U_2(t)\otimes\cdots\otimes U_N(t),
\end{equation}
where the local transformations $U_1(t),U_2(t),\cdots, U_N(t)$ are all $2\times 2$ unitary matrices in ${\rm SU}(2)$. They obey the following Schr\"{o}dinger equations
\begin{equation}
\label{eq:spin1equation}\dot{U}_k(t)= -i\left[H_{\rm c}(t)-\delta_kH_{\rm d}(t)\right]U_k(t),
\end{equation}
for $k=1,2\cdots,N$, where
\begin{eqnarray}
H_{\rm c}(t) &=& (\omega_{\rm rf}-\omega_0)\sigma_z - \omega_x(t)\sigma_x - \omega_y(t)\sigma_y, \\
H_{\rm d}(t) &=& - \omega_0\sigma_z - \omega_x(t)\sigma_x -\omega_y(t)\sigma_y.
\end{eqnarray}
In particular, when $\omega_{\rm rf}=\omega_0$, we have
\begin{eqnarray}
H_{\rm c}(t) &=&  -\omega_x(t)\sigma_x -\omega_y(t)\sigma_y.
\end{eqnarray}

Eq.~(\ref{eq:spin1equation}) shows that the homonuclear spins are dynamically differentiated by the Hamiltonians $\delta_{k}H_{\rm d}(t)(k=1,2,\cdots,N)$. To analyze their differences during evolution, we pick the case of two spins and denote by $V(t) = U_1^\dag(t)U_2(t)$ the relative motion of spin 2 with respect to spin 1. The dynamics of two-spin homonuclear systems can thus be equivalently described as
\begin{eqnarray} \label{eq:MotionEquationOfSpin1}
\dot {U}_1(t)\!\!\! &=& \!\!\! -i\left[H_{\rm c}(t)-\delta_1H_{\rm d}(t)\right]U_1(t),\\
\label{eq:RelativeMotionEquation}
\dot V(t) \!\!\! &=& \!\!\! -i(\delta_1-\delta_2)\left[U_1^\dag(t) H_{\rm d}(t) U_1(t)\right] V(t),
\end{eqnarray}
in which $U_1(t)$ can be driven much faster than $V(t)$ when the available control intensity $\Omega$ is far greater than the frequency difference $|\delta_1- \delta_2|\omega_0$. Therefore, the minimum time needed to implement the transformation is mainly determined by the slow motion $V(t)$ from $V(0)=I_2$ to $V(T) = U_{1f}^\dag U_{2f}$, where $U_{1f}$ and $U_{2f}$ are the desired operations on spins 1 and 2, respectively.

In liquid-state NMR, the Larmor frequency $\omega_0$ (about several hundreds of megahertz) is far greater than the control bound $\Omega$ (about tens of kilohertz). So, $H_{\rm d}(t)$ is dominated by its constant part $-\omega_0 \sigma_z$, implying that $V(t)$ evolves at an approximately constant speed, but its direction can be changed by the controls $\omega_x(t)$ and $\omega_y(t)$. The time spent for $V(t)$ to go from $V(0)=I_2$ to $V(T) = U_{1f}^\dag U_{2f}$ is thus proportional to the distance travelled in ${\rm SU}(2)$. Therefore, an ideal minimum-time trajectory of $V(t)$ must be along the the geodesic curve (i.e., the shortest curve) in ${\rm SU}(2)$ that connects $V(0)$ and $V(T)$, but in fact it is slightly longer than the geodesic distance due to the limited control power.

This observation indicates that the minimal time can be approximated as the quotient of the geodesic distance and the speed of $V(t)$. The calculation requires a right-invariant Riemanian metric on ${\rm SU}(2)$ defined as follows:
$$\langle X_1V,X_2V\rangle=\tr(X_1^\dag X_2),$$
where $V\in{\rm SU}(2)$ and $X_{1,2}$ are skew-Hermitian matrices. The geodesic curve accompanied with this metric is a one-parameter unitary group $G(s)=e^{sX}$ with $G(0)=I_2$ and $G(1)=U_{1f}^\dag U_{2f}$. This implies that $X=\log(U_{1f}^\dag U_{2f})$, and the path length from $V(0)$ to $V(T)$ is
\begin{eqnarray*}
  L_{\rm geodesic} &=& \int_0^1{\rm d}s \sqrt{\langle XG(s),XG(s)\rangle} \\
   &=& \int_0^1{\rm d}s \sqrt{\tr(X^\dag X)}=\|\log(U_{1f}^\dag U_{2f})\|_F,
\end{eqnarray*}
where $\|X\|_F=\sqrt{\tr(X^\dag X)}$ is the Frobenius norm of $X$. Similarly, the actual path length of $V(t)$ from $t=0$ to $t=T$ is
\begin{eqnarray*}
  L &=& \int_0^{T_{\rm minimum}}{\rm d}t \|-i(\delta_1-\delta_2)U_1^\dag(t) H_{\rm d}(t) U_1(t)\|_F \\
   &\approx& \int_0^{T_{\rm minimum}}{\rm d}t \|-i(\delta_1-\delta_2)\omega_0 U_1^\dag(t) \sigma_z U_1(t)\|_F \\
      &=& \frac{|\delta_1-\delta_2|\omega_0}{\sqrt{2}}T_{\rm minimum} .
\end{eqnarray*}
As analyzed above, the actual path length $L$ should be slightly longer, but very close to, the geodesic distance from $V(0)$ to $V(T)$ as long as the following assumption
\begin{equation}\label{eq:Assumption}
\Omega\gg |\delta_1- \delta_2|\omega_0\gg J
\end{equation}
is satisfied. This leads to the following estimation formula:
\begin{eqnarray} \label{eq:stiFormula}
T_{\rm minimum}\gtrsim T_{\rm geodesic}=\frac{ \sqrt{2}}{|\delta_1- \delta_2|\omega_0}\| \log(U_{1f}^{\dag} U_{2f})\|_F
\end{eqnarray}
that is to be used in the optimizations.
\begin{figure}[ht]
\begin{center}
  \includegraphics[width=.9\columnwidth]{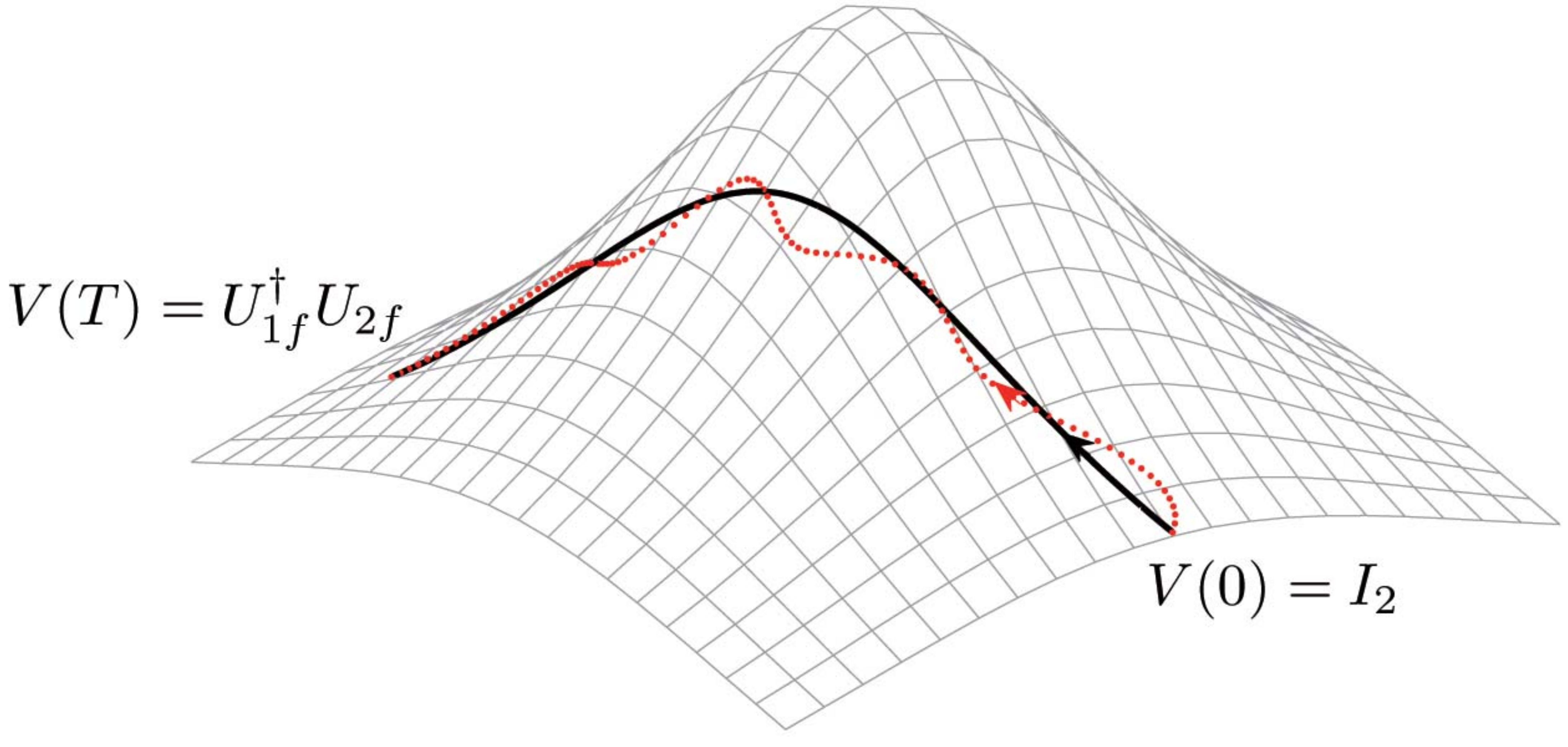}\\
    \caption{A schematic illustration on the geodesic and actual paths of $V(t)$ in ${\rm SU}(2)$ from $V(0)$ to $V(T)$. The black solid line represents the geodesic curve, and the red dash line represents the actual minimum-time trajectory that is close to the geodesic curve.
    }\label{fig:GeodesicCurve}
\end{center}
\end{figure}

\section{Numerical and Experimental Results}   \label{Sec:SimExpRes}
In this section, we will implement the gradient algorithm to seeking minimum-time control sequences near the above estimated minimal time duration. Its effectiveness will be demonstrated by both numerical simulations and experiments with the molecule of trichloroethylene ($ \rm C_2H_1Cl_3$).

\subsection{Algorithm Design}
To search a minimum-time control that achieves a local transformation $U_f=U_{1f}\otimes U_{2f}\in {\rm SU}(2)\otimes{\rm SU}(2)$, we choose gradient-type algorithms to maximize the gate fidelity
\begin{equation} \label{eq:PerformanceFunction}
\Phi =2^{-N} \Re\left\{{\rm Tr} \left[U_{f}^{\dag}U(T)\right]\right\},
\end{equation}
where $\Re(\cdot)$ returns the real part of a complex number. The optimization should attain a high fidelity above some prescribed threshold $\Phi_0$ (e.g., $\Phi_0= 0.9999$) and the final time $T$ should be as short as possible.

In numerical simulations, the control pulses $\omega_{x,y}(t)$ are digitized to a sequence $\omega_{x,y}(j)$ at $M$ time steps $j=\Delta t,2\Delta t,\cdots,M\Delta t$, where $\Delta t=T/M$. In practice, we fix $\Delta t$ and vary $M$ to find the desired minimum time. The unitary propagator of the overall system is
\begin{equation}
U(T) = U_MU_{M-1}\cdots U_2U_1,
\end{equation}
where
\begin{eqnarray*}
U_j&=&\exp\left\{ -i\Delta t\left[ H_{\rm J}+H_{\rm Z}-\omega_x(j)\sum_{k=1}^N(1-\delta_k)S_x^k\right.\right.\\
&&\left.\left.-\omega_y(j)\sum_{k=1}^N(1-\delta_k)S_y^k \right] \right\}.
\end{eqnarray*}
Note that the coupling Hamiltonian $H_J$ is omitted for estimation in Section \ref{Sec:MinimumControlForTwoQubit}, but in numerical simulation we need to keep it in the calculation for high precision.

Using the first-order Taylor expansion of $U_j$, the gradient of the fidelity function, Eq.~(\ref{eq:PerformanceFunction}), can be approximately evaluated as (see derivation in \cite{Khaneja2005})
\begin{eqnarray*}
&&\frac{\partial \Phi}{\partial \omega_{x,y}(j)}\\
 &= &2^{-N} \Re\left\{{\rm Tr} \left[U_{f}^{\dag}U_M\cdots \frac{\partial U_j}{\partial \omega_{x,y}(j)}\cdots U_2U_1  \right]\right\} \\
&= &\frac{\Delta t}{2^{N}} \sum_{k=1}^N (1-\delta_k)\Im\big\{{\rm Tr} \big[U_{f}^{\dag}U_M\cdots S^k_{x,y} U_j \cdots U_2U_1 \big]\big\}.
\end{eqnarray*}
There are many choices of gradient search algorithms, among which we choose the bounded BFGS algorithm that can deal with the bound limitation on controls (see Appendix A). In addition, the algorithm also attempts to improve the smoothness and the robustness of the resulting control sequence, the discussion of which can be found in Appendices B and C, respectively.

Besides the above algorithmic considerations, a key problem is the determination of minimal time $T_{\rm minimum}$ for the gradient algorithm to climb. Using the estimation formula derived in Section \ref{Sec:MinimumControlForTwoQubit}, we start from the tight lower bound $T=T_{\rm geodesic}$ on the minimum-time $T_{\rm minimum}$. Next, let $\Delta T$ be the minimal time required for single-spin operations $U_{1f}$ and $U_{2f}$ \cite{boozer2012time}. we increase $T$ by $\Delta T$ until the threshold $\Phi_0$ is reached. This is because, as shown in Fig.~\ref{fig:GeodesicCurve}, the closeness of the actual trajectory of $V(t)$ to the geodesic curve depends on how fast the single spins can evolve. In such way, we can find a tight upper bound with which the search for $T_{\rm minimum}$ can be greatly narrowed down. If necessary, a bisection procedure can be conducted to determine the exact value of $T_{\rm minimum}$ between its lower and upper bounds.

To summarize, the algorithm for implementing two-spin minimum-time local transformations is as follows:
\begin{enumerate}
\item Calculate the geodesic time $T_{\rm geodesic}$ and $\Delta T$ with given system parameters for given target transformations
and start from $T_0 =T_{\rm geodesic}$ as a lower bound on $T_{\rm minimum}$.
\item Find an upper bound $T_{\rm ub}^\ast$ of the minimum time:
\begin{enumerate}
  \item[2.1)] Optimize the control sequence $u^k_{x,y}(j)$ with time duration $T_{k}$ using the gradient algorithm.
  \item[2.2)] Set $T_{k+1}= T_{k}+ \Delta T$ and go to Step 2.1) until $\Phi\geq \Phi_0$.
\end{enumerate}
\item Search the minimum-time control $u^\ast_{x,y}(j)$ by the method of bisection over the interval
$\left[T^0_{\rm lb},T^0_{\rm ub}\right]$, where $T^0_{\rm lb}=T_{\rm ub}^\ast-\Delta T ,T^0_{\rm ub}=T_{\rm ub}^\ast$:
\begin{enumerate}
  \item[3.1)] Optimize the control sequence $u^k_{x,y}(j)$ with time duration $T_{k} = (T_{\rm lb}^k+T^k_{\rm ub})/2$.
  \item[3.2)] Set $T^{k+1}_{\rm lb}=T^k_{\rm lb}, T^{k+1}_{\rm ub}=T_k$ if $\Phi\geq \Phi_0$ can be achieved. Otherwise, set $T_{\rm lb}^{k+1}=T_k, T_{\rm ub}^{k+1}=T_{\rm ub}^k$.
  \item[3.3)] Repeat Steps 3.1) and 3.2) until $T_{\rm lb}^{k+1}=T_{\rm ub}^{k+1}$.
\end{enumerate}
\item Smooth and re-optimize the control sequence iteratively (see Appendix for details) until an the experimentally-friendly minimum-time control sequence is yielded.
\end{enumerate}

Note that any local optimization algorithm (typically, the gradient algorithm) can be trapped by local maxima. Nevertheless, as analyzed in a series of papers on the topological analysis of quantum optimal control landscapes \cite{ho2006effective,Wu2008,hsieh2008optimal,Wu2012}, the transformation control problem is devoid of traps as long as the system is controllable and the time duration is sufficiently long. In the following simulations, we encounter no traps in our numerical simulations, which is consistent with this prediction.

\subsection{Numerical Results} \label{Sec:SimRes}

\begin{figure}
\centering
\includegraphics[width=.4\columnwidth]{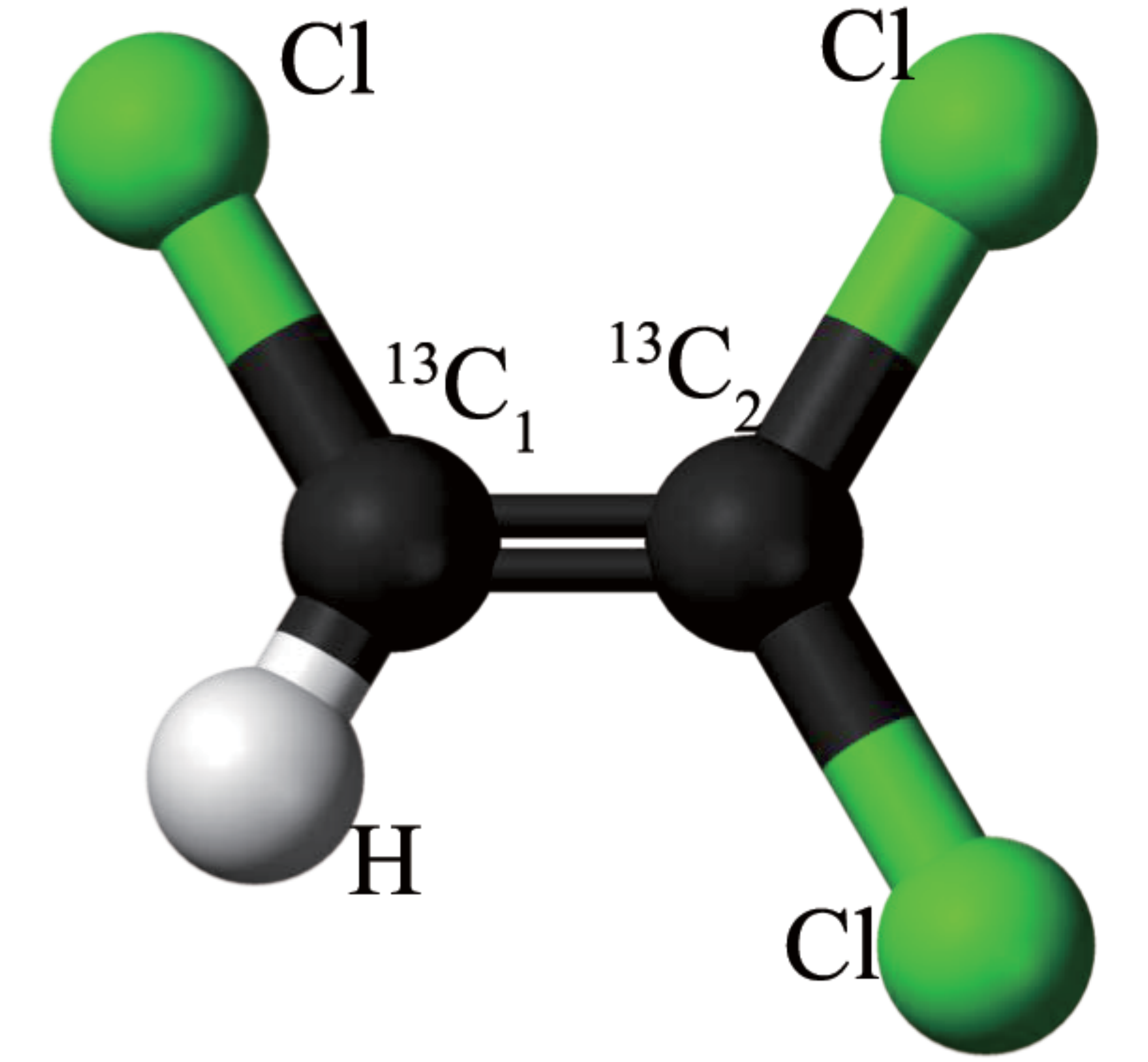}
\caption{ The 3D structure of Trichloroethylene molecule.}
\label{fig:Trichloroethylene}
\end{figure}

To demonstrate the effectiveness of the designed algorithm, we select the molecule of trichloroethylene ($ \rm C_2H_1Cl_3$) that contains two homonuclear carbon spins $\rm C_1$ and $\rm C_2$, whose 3D structure is shown in Fig.~\ref{fig:Trichloroethylene}. Their interaction with the chlorine and proton spins can be ignored or decoupled and hence is not considered in the simulations. The frequency shifts of the two carbon spins on a Bruker Avance-400 spectrometer are  $\delta_1\omega_0/2\pi=11930.18$Hz and $\delta_2\omega_0/2\pi= 11202.80$Hz, respectively. The $J$-coupling constant $J_{\rm C_{1}C_{2}}=103.49$Hz. The control bound is $\Omega/2\pi=12.50$kHz.

Take the target transformation $I_2\otimes R_z(\frac{\pi}{2})$ for example, which rotates $\rm C_2$ around $z$ axis by $90^{\circ}$ and leaves $\rm C_1$ unchanged at the final time $T$.
The frequencies of $\rm C_1$ and $\rm C_2$ are 727.38Hz apart, which is much smaller than the control bound $\Omega$. From Eq.~(\ref{eq:stiFormula}), $T_{\rm geodesic} = 344 \mu s$ is calculated and set to be the initial guess on the minimum time. Under the control bound $\Omega=12.50$kHz and time-step length $\Delta t=1\mu$s, the minimum times for single-spin operations are $T_{\rm op}^1 = 0\mu s$ for $U_{1f}=I_2$ on $\rm C_1$ and $T_{\rm op}^2 = 10\mu s$ for $U_{2f}=R_z(\frac{\pi}{2})$ on $C_2$. Therefore, $\Delta T=10\mu$s is chosen in the simulation.

After the optimization, the minimum time is found to be $352\mu s$ with fidelity above $\Phi_0=0.9999$.
Note that the assumption (\ref{eq:Assumption}) is only loosely satisfied because $|\delta_1-\delta_2|\omega_0$ is not far greater than $J_{\rm C_1C_2}$, but our formula still provide a rather good estimation that
is only 8$\mu s$ shorter. We also tested another 11 local quantum transformations under the same field constraint and chemical shifts, whose optimization results are listed in Table~\ref{tab:GatesTime}. It can be seen that $T_{\rm geodesic}$ is very close to $T_{\rm minimum}$ in all cases.

\begin{table}[ht]
\begin{center}
\caption{ Numerical simulation results for homonuclear carbon spins ($\rm C_1-C_2$) in trichloroethylene.}\label{tab:GatesTime}.
\begin{tabular}{c|c|c}
\hline
\hline
Target Transformation  & $T_{\rm geodesic} $& $T_{\rm minimum}$  \\
     & $ (\mu {\rm s})$ &  $(\mu {\rm s})$  \\
\hline
$I_2\otimes R_x(\frac{\pi}{2})$ &344	&359  \\
\hline
$I_2\otimes R_y(\frac{\pi}{2})$ &344 	&356 \\
\hline
$I_2\otimes R_z(\frac{\pi}{2})$ &344 	&352 \\
\hline
$R_x(\frac{\pi}{2})\otimes I_2$ &344 	&356 \\
\hline
$R_y(\frac{\pi}{2})\otimes I_2$ &344  	&356 \\
\hline
$R_z(\frac{\pi}{2})\otimes I_2$ &344 	&352 \\
\hline
$R_x(\frac{\pi}{2})\otimes R_y(\frac{\pi}{2})$ &459   &476 \\
\hline
$R_x(\frac{\pi}{2})\otimes R_z(\frac{\pi}{2})$ &459 &467 \\
\hline
$R_y(\frac{\pi}{2})\otimes R_x(\frac{\pi}{2})$ &459 &476 \\
\hline
$R_y(\frac{\pi}{2})\otimes R_z(\frac{\pi}{2})$ &459 &468 \\
\hline
$R_z(\frac{\pi}{2})\otimes R_x(\frac{\pi}{2})$ &459 &466 \\
\hline
$R_z(\frac{\pi}{2})\otimes R_y(\frac{\pi}{2})$ &459 &466 \\
\hline
\hline
\end{tabular}
\end{center}
\end{table}

\begin{figure}
\centering
\includegraphics[width=.9\columnwidth]{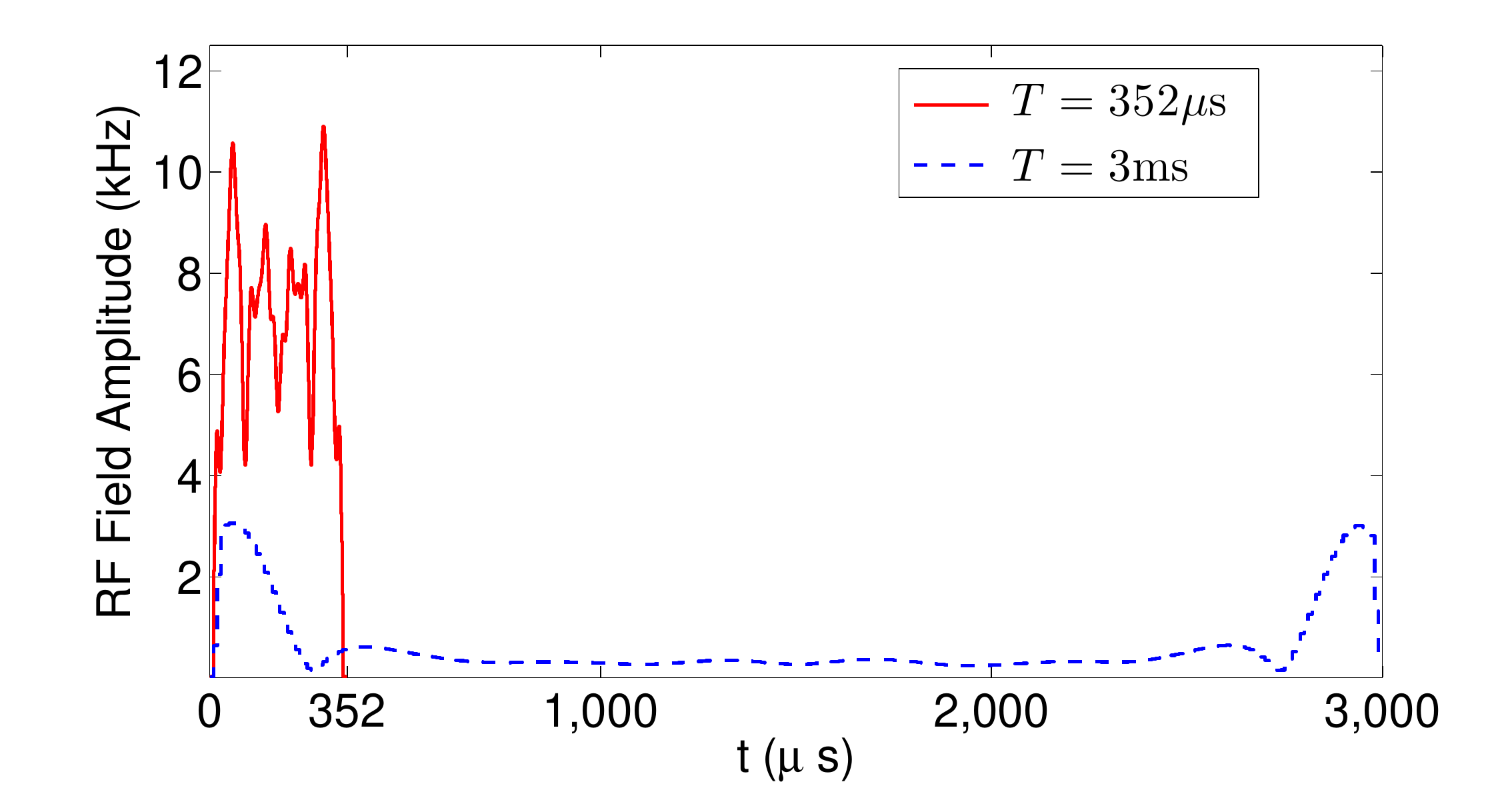}
\caption{ The optimal control amplitudes $\omega_r(t)$ for the transformation $I_2\otimes R_z(\frac{\pi}{2})$. The time durations are $352\mu s$ and 3ms, respectively. }
\label{fig:ControlProfile}
\end{figure}

\begin{figure}
\centering
\subfigure[ $\ T=352\mu \rm s$]{ \label{figa:bloch_IZ_short}\includegraphics[width=0.7\columnwidth]{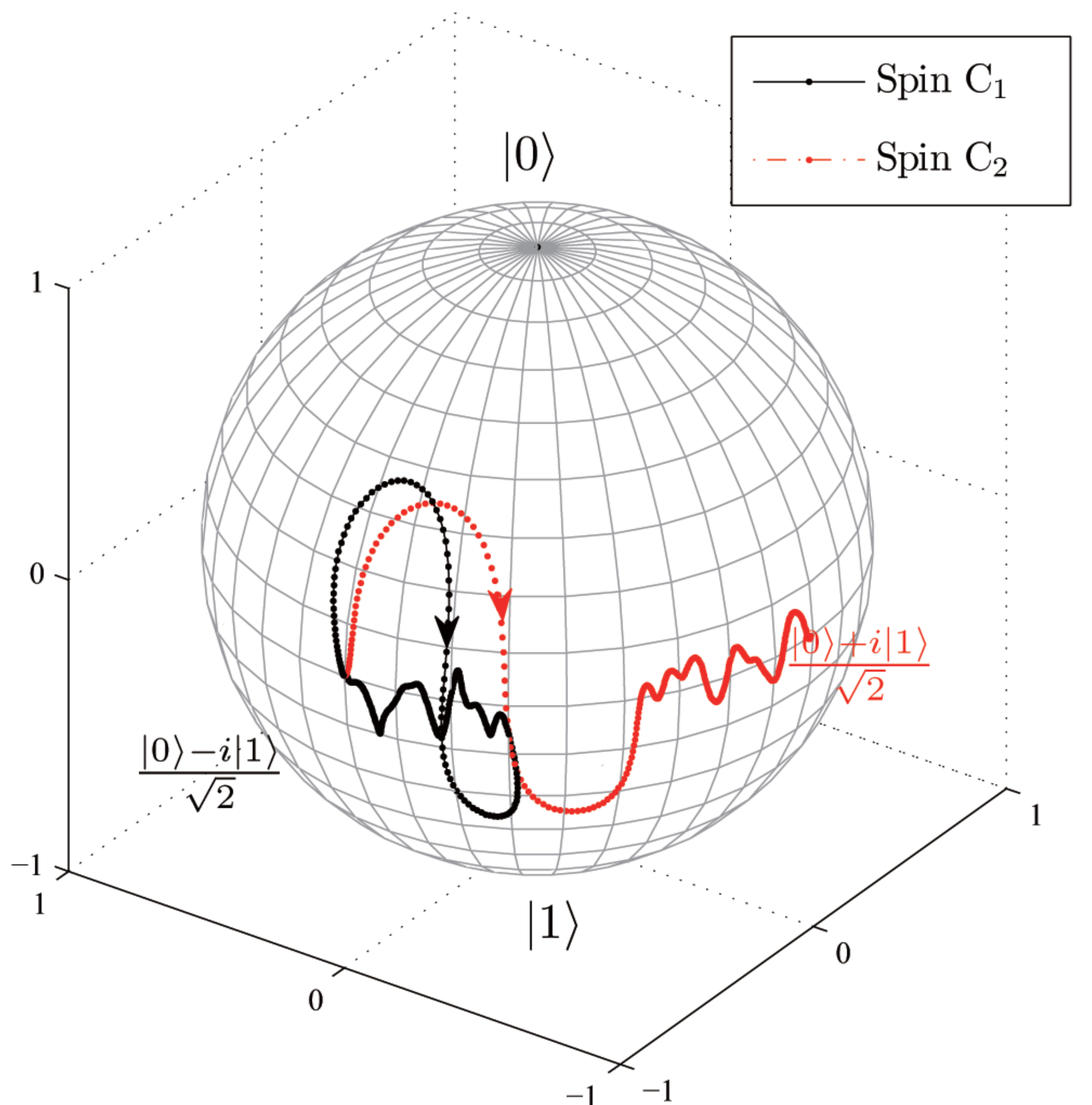}}
\subfigure[ $\ T=3\rm ms$]{ \label{figb:bloch_IZ_long}\includegraphics[width=0.7\columnwidth]{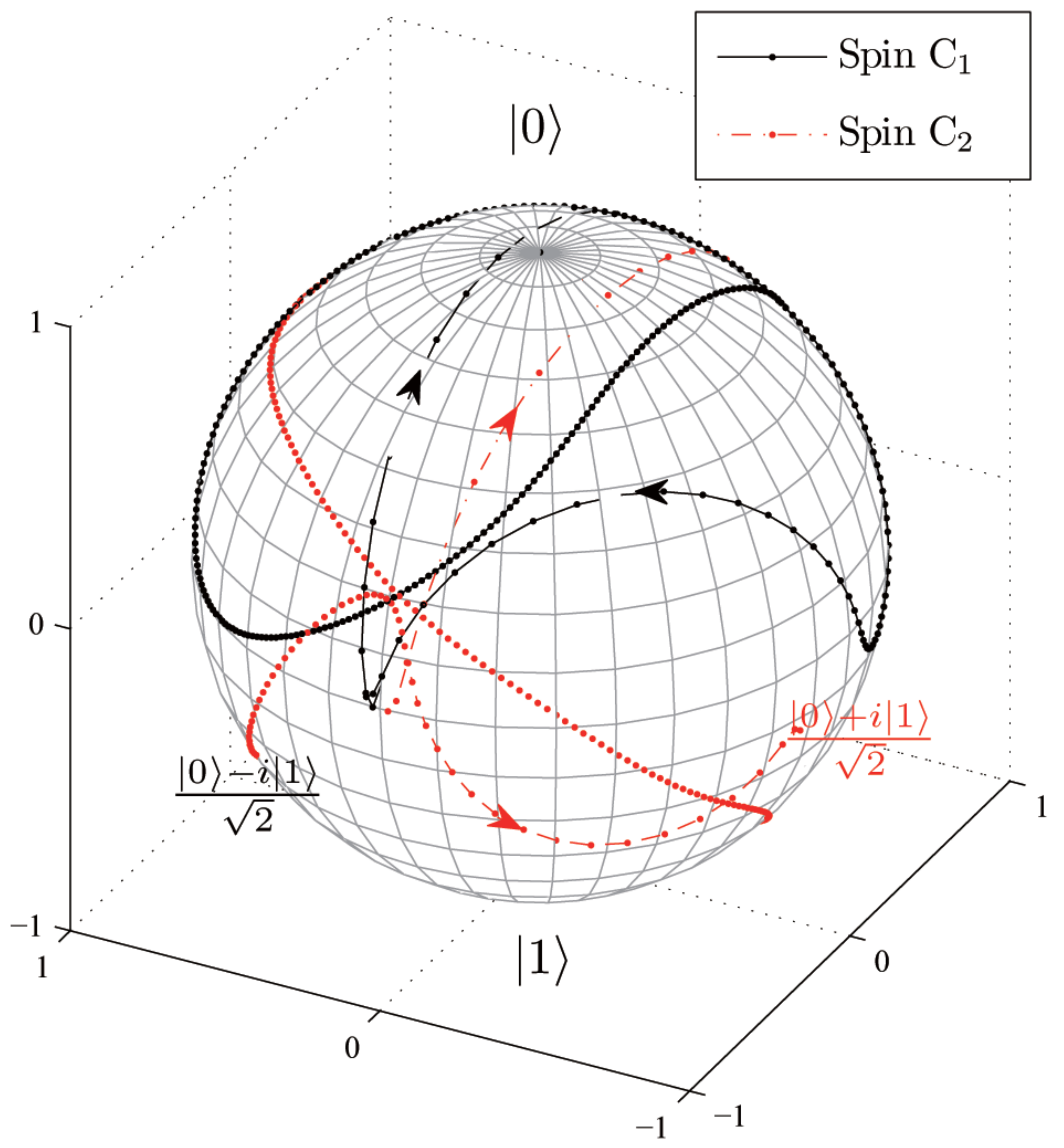}}\\
\caption{
 The Bloch-sphere trajectories of spin 1 and spin 2 driven by the optimal controls shown in Fig.\ref{fig:ControlProfile}. Starting from the same initial state $\frac{|0\rangle-i|1\rangle}{\sqrt{2}}$, the controls  rotate $\rm C_2$ around $z$ axis by $90^{\circ}$ , but pull $\rm C_1$ back to the initial state, which realizes the transformation $I_2\otimes R_z(\frac{\pi}{2})$. The state evolves slowly when the marker is dense.
}\label{fig:CurveOnBloch}
\end{figure}

For comparison, we optimize the control sequence over a much longer time interval with $T=3$ms, which is typical in NMR experiments without optimization. As shown in Fig.~\ref{fig:ControlProfile}, the $352 \mu \rm s$ control is maintained at a much higher RF power level than the 3ms pulse, which features the bang-bang property of time optimal controls. Figure \ref{fig:CurveOnBloch} displays the control guided trajectories of the spin states on the Bloch sphere. Because the relative motion $V(t)$ of the two spins follows a geodesic curve, the spins travel much shorter distances under the 352$\mu$s control than under the 3ms control, and the 352$\mu$s control spends most time on the separation of two homonuclear spins. These observations are consistent with our analysis in Section \ref{Sec:MinimumControlForTwoQubit} .

\begin{figure}
\centering
\subfigure[]{ \label{figa:EstiAmp}\includegraphics[width=0.9\columnwidth]{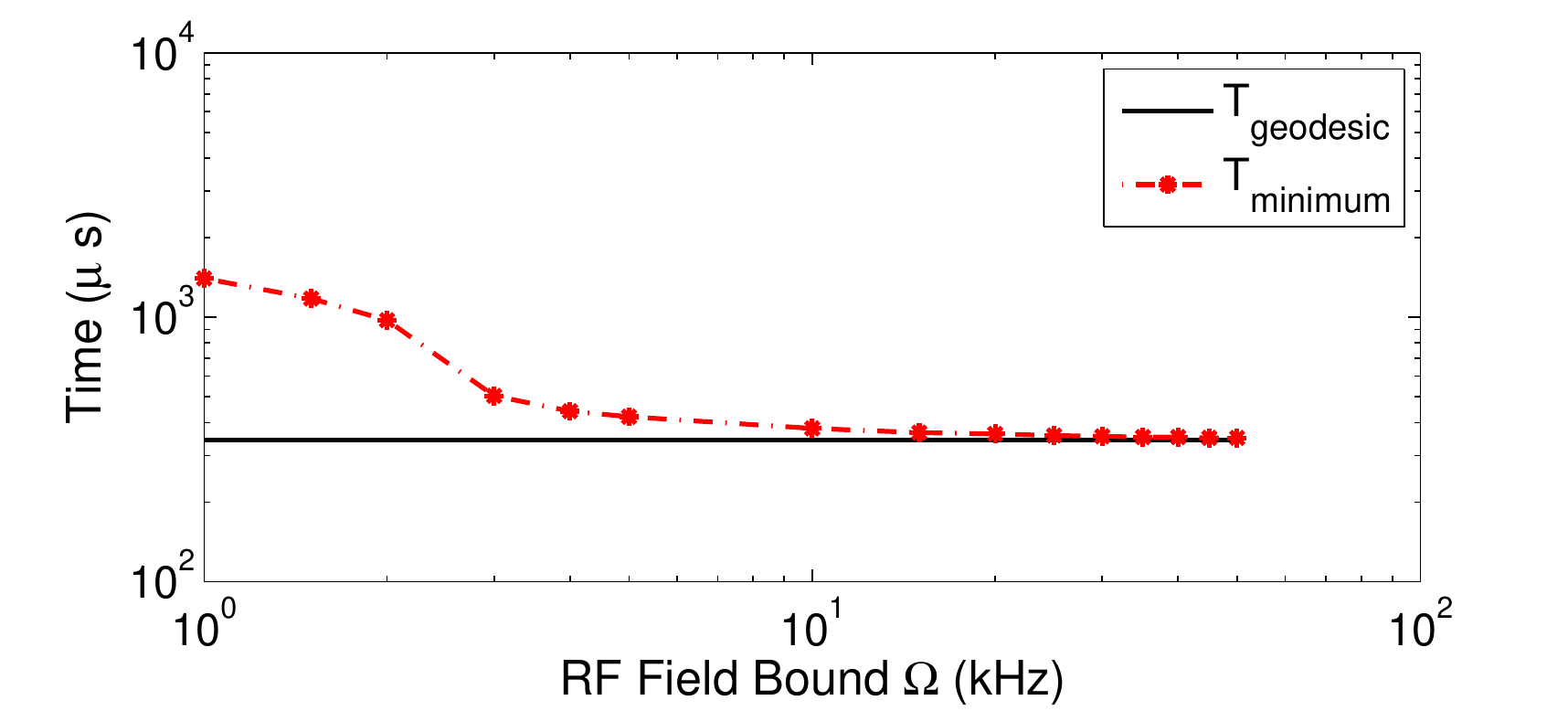}}
\subfigure[]{ \label{figb:EstiShift}\includegraphics[width=0.9\columnwidth]{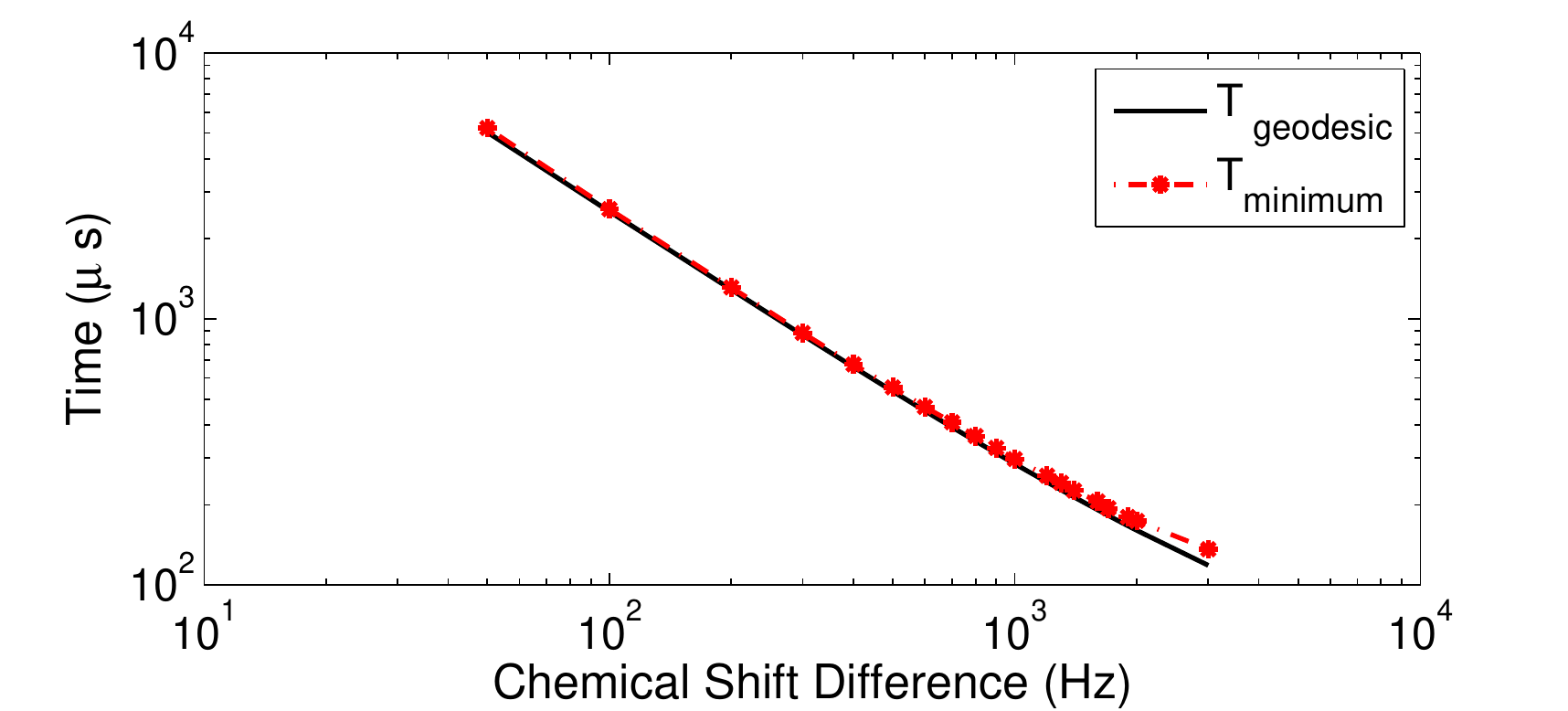}}
\caption{The comparison between the estimated minimum time and the actual minimum time found by numerical simulations for the transformation $I_2\otimes R_z(\frac{\pi}{2})$: (a) the comparison under different bounds $\Omega$ on the RF field; (b) the comparison under different frequency differences between the homonuclear spins.
}\label{fig:EstiCurve}
\end{figure}

To understand how accurate the estimation (\ref{eq:stiFormula}) could be, we numerically calculated the minimum time under different values of the control bound $\Omega$ and the frequency difference $(\delta_1-\delta_2)\omega_0$, and investigate how close $T_{\rm geodesic}$ is to the actual minimum time $T_{\rm minimum}$. Figure \ref{figa:EstiAmp} shows that their difference decreases under stronger control fields because $V(t)$ may be forced closer to the geodesic curve, while Fig.~\ref{figb:EstiShift} shows the estimation is more accurate when the two homonuclear spins are spectrally closer to each other. Thereby, our estimation formula is particularly useful for hard cases where the chemical shifts are very small.

\subsection{Experimental Results} \label{Sec:ExpRes}
The control sequences obtained in the above numerical simulations were experimentally applied to the sample of trichloroethylene ($\rm  C_2H_1Cl_3$) on a Bruker Avance-400 spectrometer. Three target transformations, $I_2\otimes R_z(\frac{\pi}{2})$, $I_2\otimes R_x(\frac{\pi}{2})$ and $R_x(\frac{\pi}{2})\otimes R_z(\frac{\pi}{2})$, were selected, and their minimal time control duration are 352$\mu$s, 359$\mu$s, and 467$\mu$s, respectively, as shown in Tab.~\ref{tab:GatesTime}.
\begin{figure}
\centering
\subfigure[]{ \label{figa:Exp_Sim_Kai_IZ}\includegraphics[width=1\columnwidth]{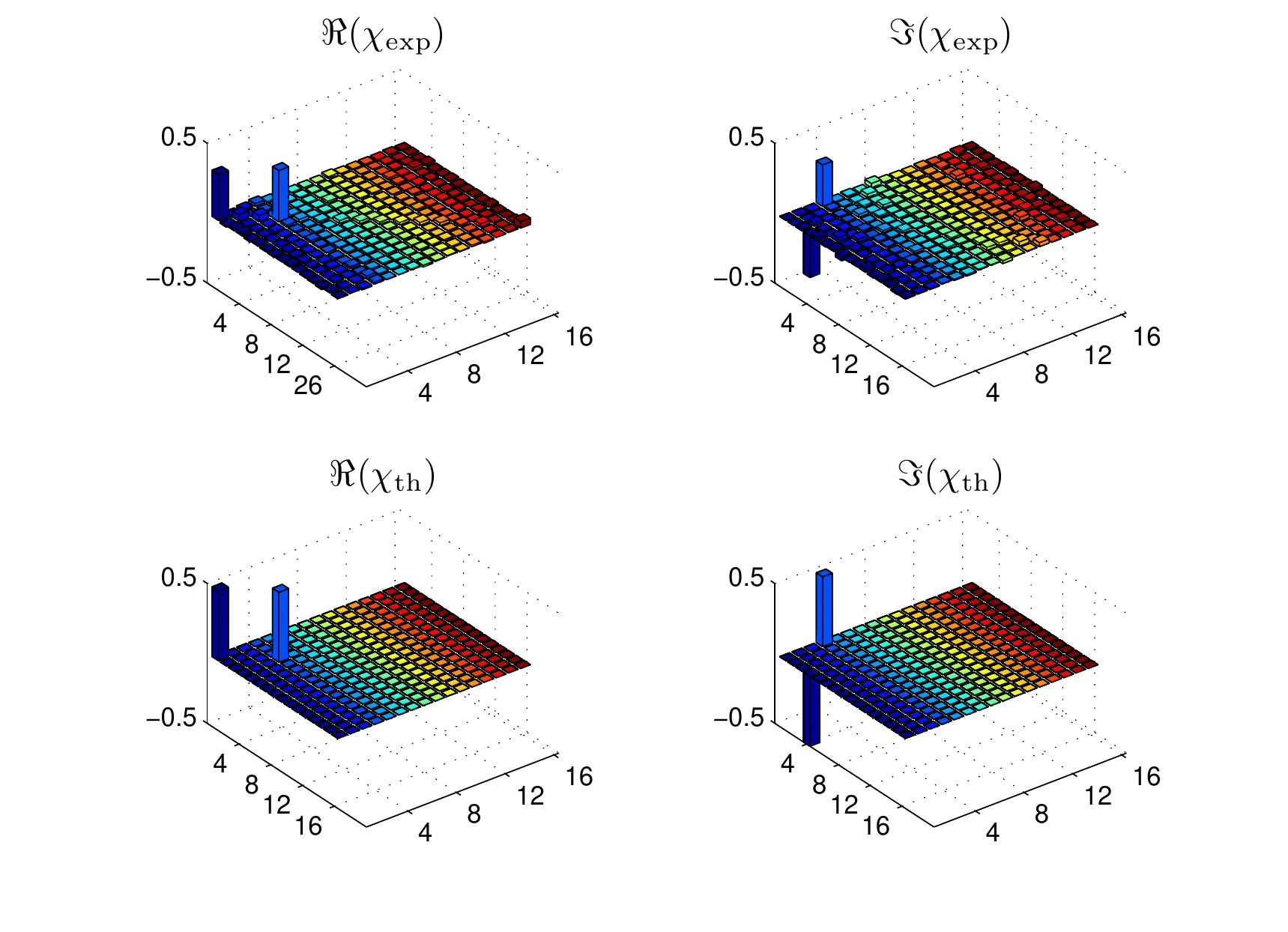}}\\
\subfigure[]{ \label{figb:Exp_Sim_Kai_IX}\includegraphics[width=1\columnwidth]{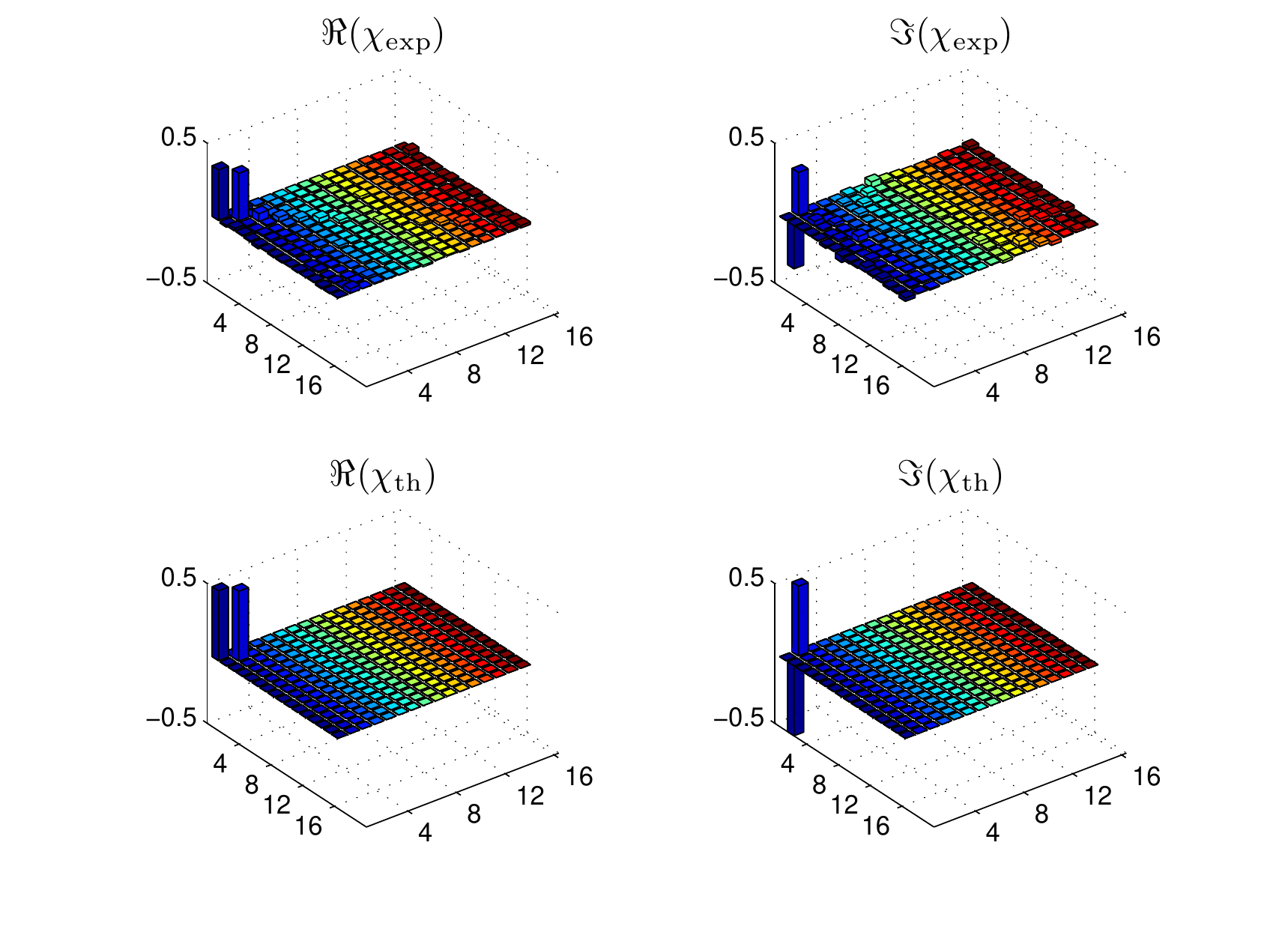}}\\
\subfigure[]{ \label{figc:Exp_Sim_Kai_XZ}\includegraphics[width=1\columnwidth]{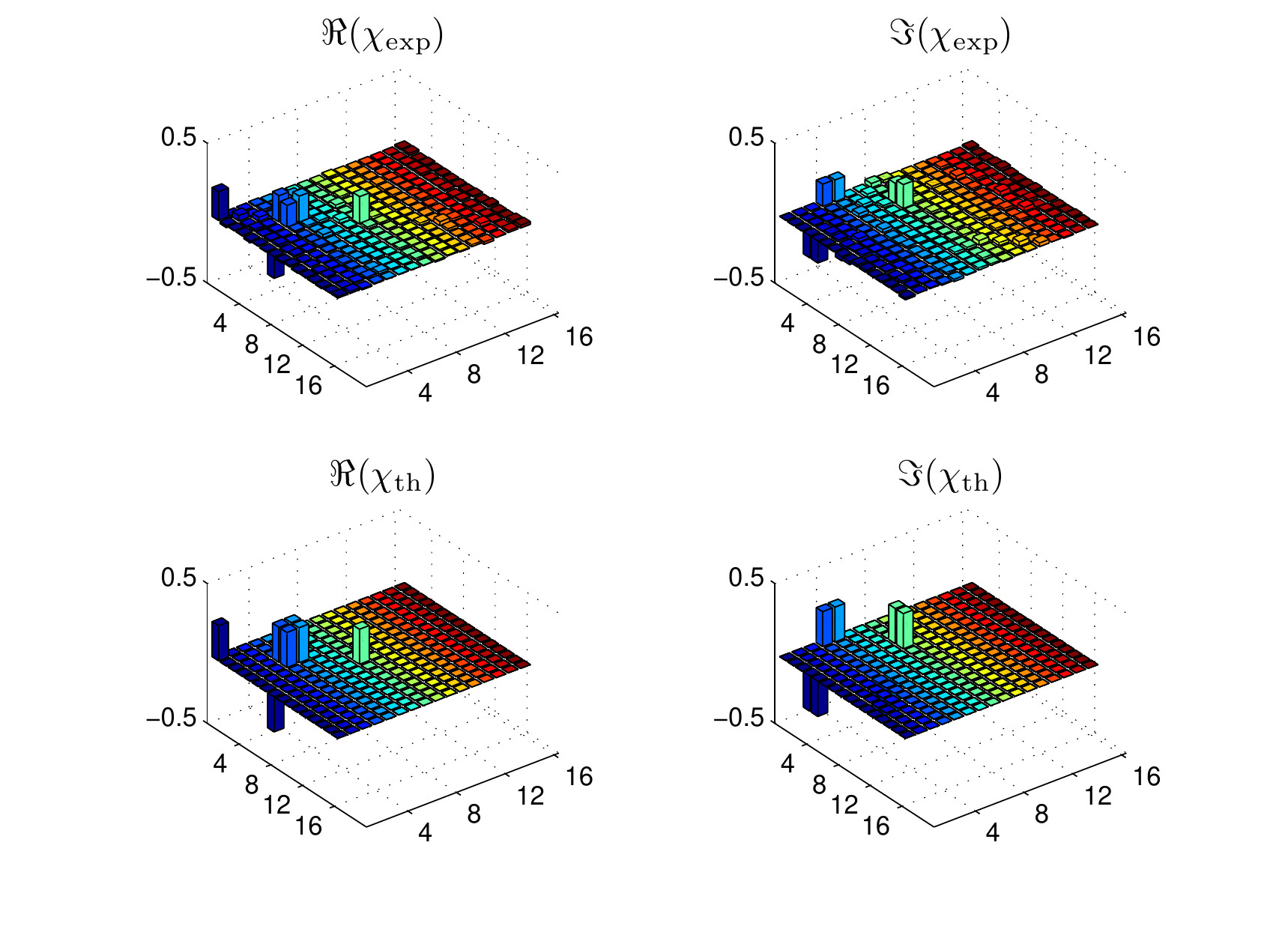}}
\caption{The real and imaginary parts of $\chi$ matrices for transformations (a) $I_2\otimes R_z(\frac{\pi}{2})$, (b) $I_2\otimes R_x(\frac{\pi}{2})$ and (c) $R_x(\frac{\pi}{2})\otimes R_z(\frac{\pi}{2})$. $\chi_{\rm th}$ is the theoretical value, and $\chi_{\rm exp}$ was experimentally constructed by QPT. The labels 1$\sim$16 in the $x$ and $y$ axes correspond to the operator basis (\ref{eq:OperatorSet}).
}\label{fig:Exp_Sim_Kai}
\end{figure}

To evaluate the performance of experimental controls, quantum process tomography (QPT) is used to reconstruct the process matrix $\chi$ for describing the actual operation achieved in laboratory. The process matrix $\chi$ is defined as the mapping from the initial density matrix $\rm \rho(0)$ to the final-time density matrix $\rho(T)$, which is $16\times 16$ dimensional under the following matrix basis for density matrices:
\begin{equation} \label{eq:OperatorSet}
\begin{array}{cccc}
\!\!I_2\otimes I_2,& I_2\otimes\sigma_x,&-iI_2\otimes\sigma_y,&I_2\otimes\sigma_z, \\
\!\!\sigma_x\otimes I_2,&\sigma_x\otimes\sigma_x,&-i\sigma_x\otimes\sigma_y,&\sigma_x\otimes\sigma_z,\\
\!\!-i\sigma_y\otimes I_2,&-i\sigma_y\otimes\sigma_x,&-\sigma_y\otimes\sigma_y,&-i\sigma_y\otimes\sigma_z,\\
\!\!\sigma_z\otimes I_2,&\sigma_z\otimes\sigma_x,&-i\sigma_z\otimes\sigma_y,&\sigma_z\otimes\sigma_z.
\end{array}
\end{equation}
In laboratory, QPT is done by performing experiments with selected initial states and observables to be measured under the same control sequence, from which matrix elements of $\chi$ can be reconstructed one by one. This is a standard process in quantum information processing and interested readers are referred to \cite{nielsen2010quantum} for more details.

As picturized in Fig.~\ref{fig:Exp_Sim_Kai}, the $\chi$ matrix constructed from experimental data for the selected three transformations, which shows that the experimental operation is close to the predicted operation.
To quantitatively evaluate the sameness of experimental $\chi_{\rm exp}$ with theoretical $\chi_{\rm th}$, one can use the following attenuated $\chi$ fidelity \cite{weinstein2004quantum}
$$F_{\rm attenuated}=|{\rm Tr}(\chi_{\rm exp}\chi_{\rm th}^{\dag})|$$
$\chi$ fidelities to assess the performance of the transformations, which turn out to be $61.03\%$,~$62.47\%$ and~$62.71\%$ fro the three selected transformations, respectively.

To correct the error caused by an overall loss of decoherence due to nonunitary operations, one can use the following unattenuated $\chi$ fidelity \cite{weinste in2004quantum, wang2008alternative,zhang2012experimental,feng2013experimental}
$$F_{\rm unattenuated}=\frac{|{\rm Tr}(\chi_{\rm exp}\chi_{\rm th}^{\dag})|}{\sqrt{{\rm Tr}(\chi_{\rm exp}\chi_{\rm exp}^{\dag}) {\rm Tr}(\chi_{\rm th}\chi_{\rm th}^{\dag})}},$$
which are $93.90\%$, $92.67\%$ and $93.19\%$, respectively.

Noticing that the unattenuated fidelity is still far from good as those in numerical simulations ($\Phi>99.99\%$), we performed process tomography on a null computation (i.e., $U_f=I_4$, the output signal is measured without any control operation) to analyze the error source. It is found $F_{\rm unattenuated}=95.11\%$, which shows a nearly $5\%$ systematic error. They come from imperfect pulse calibration and inhomogeneity of the RF field during the preparation and readout steps for QPT. Thus, our experimental controls are pretty accurate after correcting the systematic error.

Table~\ref{tab:SLCmp} compares experimental results under the short time (minimum-time) and long time controls. It is observed that the attenuated fidelities under minimum-time controls are collectively slightly higher than those under long-time controls, which is reasonable because quantum coherence is less destroyed on a shorter time interval. However, the unattenuated fidelities under minimum-time pulses are found to be a bit lower than those under long-time pulses. Our interpretation is that the minimum-time controls are less robust as they are more sensitive to small variations in the static and control fields. These errors can be possibly reduced by more sophisticated control techniques.
\begin{table}[ht]
\begin{center}
\caption{Experimental results for control under the minimum time and long time (3$\rm ms$) RF pulses} \label{tab:SLCmp}
\begin{tabular}{c|c|c|c|c}
\hline
\hline
\multirow{2}{0.8cm}{transformation} & \multicolumn{2}{c|}{unattenuated fidelity} & \multicolumn{2}{c}{attenuated fidelity}\\
\cline{2-5}
 & short time  & long time & short time& long time\\
\hline
$I_2\otimes R_z(\frac{\pi}{2})$  &61.03\%  & 59.68\% & 93.90\% & 94.23\%\\
\hline
$I_2\otimes R_x(\frac{\pi}{2})$  &62.47\% & 60.72\% & 92.67\% & 94.21\%\\
\hline
$R_x(\frac{\pi}{2})\otimes R_z(\frac{\pi}{2})$  &62.71\% & 59.62\% & 93.19\%	 & 93.79\%\\
\hline
\hline
\end{tabular}
\end{center}
\end{table}

\section{Conclusion} \label{Sec:Conclusion}
In summary, we derived an estimation formula on the minimal time for local transformations on two homonuclear spins, based on which the search efforts for the optimal controls can be greatly reduced. We designed a gradient algorithm to quickly find minimum-time controls, and demonstrated its effectiveness by both numerical and experimental results.

In principle, the time-scale separation used in the estimation can be extended to multiple homonuclear spins. For example, taking spin 1 as the leading spin and let $V_k(t)=U^\dagger_1(t)U_k(t)$, $k=2,\cdots,N$, we can decompose the $N$-spin dynamics as follows:
\begin{eqnarray} \label{eq:MotionEquationOfNSpin}
\dot {U}_1(t)\!\!\! &=& \!\!\! -i\left[H_{\rm c}(t)-\delta_1H_{\rm d}(t)\right]U_1(t),\\
\dot V_2(t) \!\!\! &=& \!\!\! -i(\delta_1-\delta_2)\left[U_1^\dag(t) H_{\rm d}(t) U_1(t)\right] V(t),\\
&\vdots&\\
\dot V_N(t) \!\!\! &=& \!\!\! -i(\delta_1-\delta_N)\left[U_1^\dag(t) H_{\rm d}(t) U_1(t)\right] V(t),
\end{eqnarray}
where the relative motions $V_2(t),\cdots,V_N(t)$ represent the slow dynamics. However, we have not found a general analytical estimation formula because the underlying Riemannian geometry is much more complex. An even bigger challenge is the implementation of nonlocal transformations (i.e., unitary operations that cannot be decomposed as Kronecker product of $2\times 2$ unitary matrices). Future studies will be aimed at solving minimum-time control problems in systems with more than two spins for both local and nonlocal transformations.

\appendix
\subsection{Control bounds}
To deal with the constraint on the control field, we transform the control into the spherical polar coordinate
\begin{equation}
\omega_r = \sqrt{\omega_x^2+\omega_y^2}, \ \ \theta = \arctan\frac{\omega_y}{\omega_x},
\end{equation}
and the control is bounded by
\begin{equation} \label{eq:ControlBound}
0 \leq \omega_r \leq \Omega, \ \ 0 \leq \theta \leq 2\pi.
\end{equation}
After such transformation, we introduce the bounded
BFGS (Broyden-Fletcher-Goldfarb-Shanno) algorithm \cite{avriel2012nonlinear,Kelley1995,rigoni2003bounded} with fast convergence rate. When the absolute  minimum is inside the box (\ref{eq:ControlBound}), the bounded BFGS algorithm computes the full Newton step then (if needed)  performs a backtrack line search as the classical BFGS. When the absolute minimum lies outside the bounded box, the bounded BFGS searches the actual bounded minimum with a multiple projection technique (see \cite{rigoni2003bounded} for details).

\subsection{Smoothing}
As is well known in optimal control theory, minimum-time control tends to exert as much power as possible, which may lead to sharp pulse variations that are hard to generate by the NMR spectrometer. To reduce additional errors caused by such sharp variations, we smooth the resulting high-fidelity control sequence and reoptimize it, which usually takes only a few iterations to locate a high fidelity and smooth control sequence.
\subsection{Robustness}
A practical issue in the optimization is the loss of fidelity due to the inhomogeneity of the static magnetic and the error of RF fields. In numerical simulations, we demand that the controls reach the same high fidelity over a proper range of static magnetic fields $\omega_0$ and chemical shifts $\delta_{1,2}$. This is achieved by modifying the cost function so that high fidelity can uniformly yielded over the range of these parameters.

\end{document}